\documentclass[letterpaper]{article} % DO NOT CHANGE THIS
\usepackage[utf8x]{inputenc}
\usepackage[draft]{aaai25}  % DO NOT CHANGE THIS
\usepackage{times}  % DO NOT CHANGE THIS
\usepackage{helvet}  % DO NOT CHANGE THIS
\usepackage{courier}  % DO NOT CHANGE THIS
\usepackage[hyphens]{url}  % DO NOT CHANGE THIS
\usepackage{graphicx} % DO NOT CHANGE THIS
\def\UrlFont{\rm}  % DO NOT CHANGE THIS
\usepackage{natbib}  % DO NOT CHANGE THIS AND DO NOT ADD ANY OPTIONS TO IT
\usepackage{caption} % DO NOT CHANGE THIS AND DO NOT ADD ANY OPTIONS TO IT
\usepackage{algorithm}
\usepackage{algorithmic}
\usepackage{lipsum}
\usepackage{amsmath}
\usepackage[frozencache,cachedir=.]{minted}
\usepackage{graphicx}
\usepackage{tikz}
\usepackage{caption}
\usepackage{booktabs} % For professional looking tables
\usepackage[most]{tcolorbox} % Load with the 'most' library for additional features
\usepackage{xcolor} % Define and use colors

\DeclareUnicodeCharacter{03B1}{{$\alpha$}}

\definecolor{codeblue}{rgb}{0,0,1} % Blue color for highlighting

\definecolor{lightorange}{RGB}{255,229,204} % Light orange color
\definecolor{framecolor}{RGB}{255,204,153} % Light frame color
\definecolor{highlightdiff}{RGB}{0, 0, 255} % Blue color to highlight differences

\definecolor{lightorange}{RGB}{255,229,204} % Light orange color
\definecolor{framecolor}{RGB}{255,204,153} % Light frame color

\usepackage{newfloat}
\usepackage{listings}
\DeclareCaptionStyle{ruled}{labelfont=normalfont,labelsep=colon,strut=off} % DO NOT CHANGE THIS
\floatstyle{ruled}
\newfloat{listing}{tb}{lst}{}
\floatname{listing}{Listing}
\title{Variable Extraction for Model Recovery in Scientific Literature}

\author{
Chunwei Liu$^1$, Enrique Noriega$^2$, Adarsh Pyarelal$^2$, Clayton T. Morrison$^2$, Michael Cafarella$^1$ \\
$^1$MIT CSAIL \qquad $^2$University of Arizona \\
\{chunwei, michjc\}@csail.mit.edu, 
\{enoriega, adarsh, claytonm\}@arizona.edu
}

\usepackage{bibentry}
\begin{document}

\maketitle

%%
%% The "author" command and its associated commands are used to define the authors and their affiliations.
% \author{ABC}
% \affiliation{%
%   \institution{MIT}
% }
% \email{abc@mit.edu}

% \author{DEF}
% \affiliation{%
%   \institution{MIT}
% }
% \email{def@mit.edu}

\begin{abstract}
The global output of academic publications exceeds 5 million articles per year, making it difficult for humans to keep up with even a tiny fraction of scientific output. We need methods to navigate and interpret the artifacts---texts, graphs, charts, code, models, and datasets---that make up the literature. This paper evaluates various methods for extracting mathematical model variables from epidemiological studies, such as ``infection rate ($\alpha$),'' ``recovery rate ($\gamma$),'' and ``mortality rate ($\mu$).'' Variable extraction appears to be a basic task, but plays a pivotal role in recovering models from scientific literature. Once extracted, we can use these variables for automatic mathematical modeling, simulation, and replication of published results.

We introduce a benchmark dataset comprising manually-annotated variable descriptions and variable values extracted from scientific papers. Based on this dataset, we present several baseline methods for variable extraction based on Large Language Models (LLMs) and rule-based information extraction systems. Our analysis shows that LLM-based solutions perform the best. Despite the incremental benefits of combining rule-based extraction outputs with LLMs, the leap in performance attributed to the transfer-learning and instruction-tuning capabilities of LLMs themselves is far more significant. This investigation demonstrates the potential of LLMs to enhance automatic comprehension of scientific artifacts and for automatic model recovery and simulation.
\end{abstract}

% \pagestyle{\vldbpagestyle}
% \begingroup\small\noindent\raggedright\textbf{PVLDB Reference Format:}\\
% \vldbauthors. \vldbtitle. PVLDB, \vldbvolume(\vldbissue): \vldbpages, \vldbyear.\\
% \href{https://doi.org/\vldbdoi}{doi:\vldbdoi}
% \endgroup
% \begingroup
% \renewcommand\thefootnote{}\footnote{\noindent
% This work is licensed under the Creative Commons BY-NC-ND 4.0 International License. Visit \url{https://creativecommons.org/licenses/by-nc-nd/4.0/} to view a copy of this license. For any use beyond those covered by this license, obtain permission by emailing \href{mailto:info@vldb.org}{info@vldb.org}. Copyright is held by the owner/author(s). Publication rights licensed to the VLDB Endowment. \\
% \raggedright Proceedings of the VLDB Endowment, Vol. \vldbvolume, No. \vldbissue\ %
% ISSN 2150-8097. \\
% \href{https://doi.org/\vldbdoi}{doi:\vldbdoi} \\
% }\addtocounter{footnote}{-1}\endgroup
% %%% VLDB block end %%%

% %%% do not modify the following VLDB block %%
% %%% VLDB block start %%%
% \ifdefempty{\vldbavailabilityurl}{}{
% \vspace{.3cm}
% \begingroup\small\noindent\raggedright\textbf{PVLDB Artifact Availability:}\\
% The source code, data, and/or other artifacts have been made available at \url{\vldbavailabilityurl}.
% \endgroup

\section{Introduction}

{The surge in scientific publications, now exceeding five million articles annually \footnote{\url{https://wordsrated.com/number-of-academic-papers-published-per-year/}}%\cite{fivemperyear}
, represents a challenge for any individual or group seeking to comprehensively review the state of the art of any given discipline. The sheer size of the information warrants the use of automated information extraction technologies to sort through and navigate vast scientific corpora. In this work, we study the scientific literature that concerns mathematical modeling, in order to aid model recovery \cite{DBLP:journals/corr/abs-2001-07295, sharp-etal-2019-eidos,SCHAFFHAUSER2023105695}: the creation of symbolic representations of mathematical models through information extraction methods applied to the scientific literature\footnote{\url{https://www.darpa.mil/program/automating-scientific-knowledge-extraction-and-modeling}}.}

\begin{figure}
    \centering
    \fbox{\includegraphics[width=0.8\linewidth]{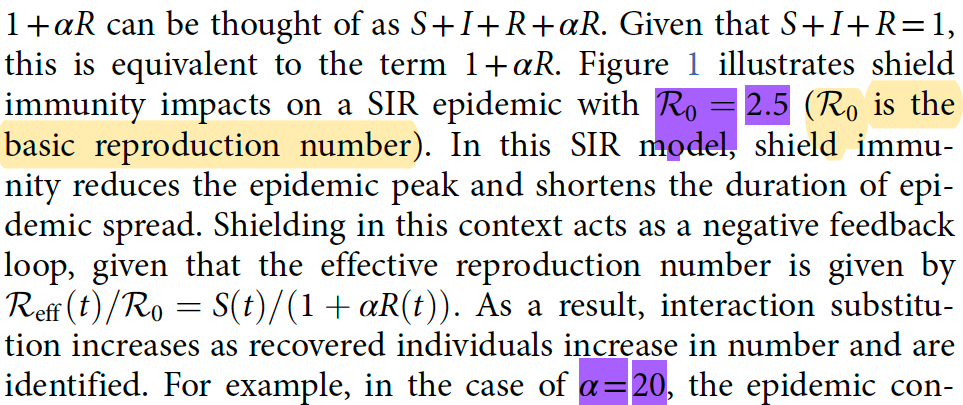}}
    
    % Arrow pointing downwards with increased thickness
\begin{tikzpicture}
    \draw[->, line width=1mm, black] (0,0) -- (0,-0.35); % Colored arrow
\end{tikzpicture}
    
    % Table with extracted variables with smaller font size
    {\small % Smaller font size for the table
    \begin{tabular}{|c|c|c|c|c|}
        \hline
        \multicolumn{2}{|c|}{Text Span} & \multicolumn{3}{c|}{Variable Extraction} \\
        \hline
        Start & End & Name & Description & Value \\
        \hline
        185 & 193 & R0 & - & 2.5 \\
        195 & 232 & R0 & Basic reproduction number & - \\
        634 & 640 & $\alpha$ & - & 20 \\
        \hline
    \end{tabular}
    }
    \caption{Example of variable extraction from a scientific paper text, illustrating the process of identifying and extracting elements such as the variable name, description, and initial value into a structured format. The figure highlights different types of extraction: variable description pairs in light orange and variable value pairs in light purple.}
    \label{fig:variable_extraction}
\end{figure}

{We introduce the task of \emph{variable extraction}: the identification and organization elements such as variable names, descriptions, and initial values into a structured format, as illustrated in Figure \ref{fig:variable_extraction}. Variable extraction is a crucial step toward model recovery as it unlocks the basic units of models presented in scientific papers. By doing so, it not only deepens the understanding of the research but also facilitates the further rebuilding and enhancement of these models.} 
% \todo{I rewrite the two paragraphs to show its importance (above) and hardness (below). Please suggest more arguments.}

The complexity of variable extraction arises from the diverse forms and locations in which variables can appear within a document. Variables may be embedded in text, figures, tables, or even scattered throughout the paper as single characters, multiple words, single values, or ranges. This variability, coupled with their interdependencies, underscores the importance and challenge of this task. Effective variable extraction is essential for identifying errors in models and converting them into executable code, which improves the accuracy and practicality of scientific research.

% Variable extraction is a fundamental step in model recovery, involving identifying and organizing key elements such as the variable name, description, and initial value into a structured format, as shown in  Figure \ref{fig:variable_extraction}. This initial phase is crucial as it allows researchers to extract the basic units of scientific models—the variables—thereby gaining a comprehensive understanding of the model. The structured format not only facilitates the systematic application of downstream analytical processes but also enhances the effectiveness and accuracy of scientific research.

% In response to this challenge, machine learning techniques have emerged as a popular solution for automating the extraction of variables from texts, a crucial step in the model recovery process. These approaches have provided researchers with tools to parse and process scientific literature more efficiently, offering a beacon of hope in the battle against information overload. However, the introduction of Large Language Models (LLMs) \cite{touvron2023llamaopenefficientfoundation,jiang2023mistral7b,openai2024gpt4technicalreport} has marked a significant change in the comprehension and extraction of knowledge from the literature. With their advanced natural language processing (NLP) capabilities, LLMs offer an unprecedented opportunity to accelerate and enhance the understanding of scientific texts, potentially revolutionizing the landscape of model recovery.

Until recently, the extraction of model variables from texts commonly employed conventional machine learning methods such as named-entity recognition \cite{tjong-kim-sang-de-meulder-2003-introduction} and relation extraction \cite{zhang2017tacred}. However, the emergence of Large Language Models (LLMs) \cite{touvron2023llamaopenefficientfoundation,jiang2023mistral7b,openai2024gpt4technicalreport}  has marked a significant change. With their enhanced natural language processing (NLP) capabilities, LLMs provide new options to enhance the efficiency and effectiveness of scientific text analysis, particularly in the extraction of variables and the broader process of model recovery.

To investigate the potential of these methods, we annotated 22 scientific papers, creating a public information extraction benchmark. This benchmark is designed to facilitate the evaluation of various variable extraction techniques. Subsequently, We then conduct a comprehensive evaluation of several LLMs designed for the variable extraction task, alongside a rule-based method and an optimized AI pipeline framework to provide additional perspective.

Our evaluation indicated that although no existing solution excels in the variable extraction task, certain configurations could significantly improve the extraction quality. The best-performing baseline model achieved an F1 score of only $0.49$ or $0.60$, depending on the evaluation metric used. However, by integrating rule-based approaches with LLMs, we enhanced performance, achieving F1 scores of up to $0.53$ and $0.64$, respectively.
% \todo{(Chunwei will update the number in the end.)} 
This integration highlights the complementary strengths of different methodologies: rule-based approaches provide additional variable extraction options from a different perspective, thus improving the performance of LLM. Overall, LLM-based solutions outperformed conventional rule-based solutions, demonstrating their capability to enhance the automatic comprehension of scientific artifacts and establish a robust foundation for automatic model recovery and simulation. 
These insights contribute to the ongoing discourse on improving the understanding and utilization of scientific literature, paving the way for more efficient and accurate scientific research in the era of information overload.
\section{Related Work}

% \todo{Information extraction}
% \todo{domain datasets}
% \todo{LLMs}
% \todo{Rule based}

% Accurately modeling natural phenomena such as pandemics, climate change, crop sciences, space weather, etc. is crucial for decision makers in high-stake positions to protect lives. The criticality of the stakes has garnered the attention government agencies to develop robust frameworks to support modeling at scale. Modeling relies on mathematical methods, and as such state of the art methods are published as scientific literature and computer software.  

% Researchers have developed AI-based tools for modeling various types of systems \cite{DBLP:journals/corr/abs-2001-07295, sharp-etal-2019-eidos,SCHAFFHAUSER2023105695}. 
% \todo{Chunwei: I moved these citations to the intro as brief references of model recovery. So maybe we can remove the above two paragraphs?}

% Within the context of scientific literature, information extraction of mathematical modeling of different natural phenomena has started to gain relevance   Accurately recovering variable descriptions and observed values from mathematical models applied to different scenarios, such as various COVID-19 outbreaks, can significantly aid modelers and decision-makers in crafting pandemic control strategies and intervention policies based on simulations and empirical evidence. 

In this work we focus on the intake of scientific literature to identify and recover the elements of mathematical models: \emph{variable descriptions} and \emph{variable values}, as described in text and we rely on NLP methods to recover them.

The field of Information Extraction (IE) is one of the main applications of NLP. It consists on identifying and extracting structured information from human-written text. These structure data, consists of named-entity recognition \cite{tjong-kim-sang-de-meulder-2003-introduction} and relation extraction \cite{zhang2017tacred}. The structured data is then leveraged by downstream applications such as building knowledge bases \cite{shimorina-etal-2022-knowledge}, slot-filling \cite{Chen2019BERTFJ}, visualization for interactions \cite{10148267}, or performing downstream inference \cite{lao-etal-2011-random}.

Large language models \cite{touvron2023llamaopenefficientfoundation,jiang2023mistral7b,openai2024gpt4technicalreport}, with their increasing versatility have become a useful tool for information extraction \cite{xu2024largelanguagemodelsgenerative}.  Building on traditional models and LLMs, numerous systems have been proposed to automatically optimize AI-powered analytics and information extraction according to user preferences \cite{zheng2023efficiently,  chen2023seed, liu2024declarative,patel2024lotus,lin2024towards}.  This work takes inspiration from these methods to identify and extract variable information.

% The transfer learning \cite{ruder-etal-2019-transfer} and instruction following capabilities \cite{NEURIPS2022_b1efde53} induced into LLMs during their training with large and varied corpora allow them to be used in different domains via prompt-engineering or in-context learning \cite{dong2024surveyincontextlearning} without the need for additional fine-tuning.

Due to its sheer size, scientific literature is frequently the subject of IE research. Some disciplines, such as health sciences and biomedical research, have received a lot of attention due to their high potential for impact. Because of this, there exists a solid record of research activity around that has produced multiple high-quality datasets \cite{DBLP:journals/corr/abs-1902-09476,kim-etal-2013-genia,ohta-etal-2013-overview} and systems \cite{10.1093/database/bay098, neumann-etal-2019-scispacy,WANG201834} focused on clinical and medical applications. 

{Prior work at extracting mathematical elements has used various classical NLP methods. A CRF model to align mathematical expressions with their definitions \cite{jp}; a pattern-based data mining method to build mathematical ontologies from \LaTeX sources \cite{10.1145/1290144.1290162}; a NER system for abstract mathematical concepts \cite{collard-etal-2022-extracting} extracting mathematical elements from scientific text. Our work builds upon the ideas from prior research and introduces a high-quality, manually curated dataset featuring annotations of variable descriptions and values extracted from a corpus of scientific literature about COVID-19 and earth sciences. Utilizing this annotated dataset, we have comprehensively evaluated the most popular LLMs, machine learning models, and their combinations, assessing their effectiveness in identifying and extracting this critical information.}

% To the best of our knowledge, our work introduces the first high-quality, manually curated dataset featuring annotations of variable descriptions and values extracted from a corpus of scientific literature about COVID-19 and earth sciences. Utilizing this annotated dataset, we have comprehensively evaluated the most popular LLMs, machine learning models, and their combinations, assessing their effectiveness in identifying and extracting this critical information.

\section{Variable Extractions Dataset}
The benchmark comprises excerpts extracted from 22 papers that focus on pandemic research, specifically available at the benchmark repository\footnote{https://github.com/mitdbg/scivar} These papers typically introduce at least one epidemiological model, providing detailed descriptions and evaluations of the models and their variables. Collectively, the research papers address the challenge of modeling and forecasting the spread of COVID-19 under various scenarios and interventions. They explore a range of modeling approaches, including standard models like SIR and SEIR, as well as more complex frameworks such as COVID-ABS and Covasim. These models are used to analyze the effects of government interventions and to predict the trajectory of the pandemic in different regions. In addition, studies emphasize the extraction and annotation of relevant variables and parameters from the literature, with the aim of enhancing the precision and applicability of these epidemiological models in real-world scenarios.

\subsection{Human Annotation}

In our study, we meticulously annotated a set of documents to facilitate the extraction and analysis of scientific variables and their contextual data. The annotation process was designed to capture three primary types of information: (1) variable names and their descriptions, (2) variable names paired with their corresponding values, and (3) additional metadata, including model card attributes and scenario card attributes. Detailed guidelines for these annotation tasks can be found in the referenced document \cite{scivardoc}.

\subsubsection{Annotation Process}
Human annotators were tasked with identifying and marking specific elements within the text according to the following categories:

\begin{enumerate}
\item \textbf{Variables with Values:} Annotators highlighted instances where a variable was directly associated with a numerical value or a range of values. This includes cases where the variable might be implied rather than explicitly stated. For example, annotators would mark the phrase "the estimated reproduction rate in the United States was around 2.5" to capture the variable (reproduction rate) and its value (2.5).

\item \textbf{Variable Descriptions:} This task involved identifying and highlighting descriptions of variables that explain or define the variable within the context of the document. For instance, the phrase "lambda represents the infection coefficient" would be annotated to link the variable lambda with its description.

%\item \textbf{Metadata Annotation:} In addition to variable-related information, annotators were also responsible for labeling other relevant metadata that provides context about the research scenario or model used in the study. This includes extracting attributes from model cards or scenario cards that describe the assumptions, parameters, and frameworks used in the modeling studies.
\end{enumerate}

\subsubsection{Annotation Standards and Tools}
The annotation was performed using Adobe Acrobat, which allowed annotators to use different colors to distinguish between the types of annotations, as specified in the guidelines. The standards for annotation emphasized precision, instructing annotators to prioritize accuracy in identifying and marking text elements. Generous alignment standards were applied during the evaluation of the annotations, focusing on the relevance and completeness of the information captured rather than strict adherence to text boundaries.

\subsubsection{Quality Control}
To ensure the quality and consistency of the annotations, each document underwent a review process. Annotations that were missed or incorrectly marked in the initial round were identified and corrected. This iterative process helped refine the annotations and improve the overall accuracy of the data set.

\subsubsection{Post-Processing with Structured Format}

After the annotation and quality review process, each paper will have a unified color code mapping for different annotation categories. We utilize pdfannots\footnote{\url{https://github.com/0xabu/pdfannots}} tool to extract the PDF annotations into JSON format, categorizing the entries and their text spans from the original text. Pdfannots is a program that extracts annotations (highlights, comments, etc.).

{For each annotation, we obtain the highlighted text and its surrounding context into a text passage. We aggregated passages shared by multiple annotations to remove redundancy, for example, a paragraph containing several variable descriptions will appear only once in the dataset with all its associated annotations attached.}  From the 22 scientific papers, we have collected 556 text chunks containing $2083$ variable-related annotations (1236 for variable descriptions and 847 for variable values). Each text block is configured with a set of annotations, which include character index and span, text extraction, and annotation type. An example of the structured JSON output can be seen in Figure \ref{fig:example_json}.

\begin{figure}
  \begin{minted}[
    xleftmargin=10pt,
    linenos,
    fontsize=\fontsize{6}{7}\selectfont,
    breaklines,
    breakanywhere
  ]{json}
{
  "all_text": "1 + αR can be thought of as S + I + R + αR. Given that S + I + R = 1, this is equivalent to the usual form 1 + αR. Figure 1 illustrates shield immunity impacts on a SIR epidemic with (R0 = 2.5) (R0 is the basic reproduction number). In this SIR model, shield immunity reduces the epidemic peak and shortens the duration of epidemic spread. Shielding in this context acts as a negative feedback loop, given that the effective reproduction number is given by Reff(t) / R0 = S(t) / (1 + αR(t)). As a result, interaction substitution increases as recovered individuals increase in number and are identified. For example, in the case of (α = 20), the epidemic concludes with less than 20% infected in contrast to the final size of ~90% in the baseline scenario without shielding (Fig. 2).",
  "page": 2,
  "annotations": [
    [185, 193, "R0 = 2.5", "var val"],
    [195, 232, "(R0 is the basic reproduction number)", "var desc"],
    [634, 640, "α = 20", "var val"]
  ],
  "file": "epidemic_model_analysis"
}
  \end{minted}
  \caption{Example of SciVar JSON output extracted and formatted from an annotated PDF text block in Figure \ref{fig:variable_extraction}.}
  \label{fig:example_json}
\end{figure}

This post-processing step ensures that the annotations are not only accurately and automated captured but also structured in a way that facilitates further analysis and application in information extraction systems and other research tools.

\subsubsection{Contributions to Research}
The annotated dataset serves as a critical resource for developing and evaluating information extraction systems tailored for scientific literature. By providing a structured representation of variables and their contextual information, the dataset aids in the advancement of automated tools capable of supporting mathematical modeling and simulation in epidemiological research.

The meticulous annotation process, guided by comprehensive and adapted guidelines, ensures that the dataset not only supports current research needs but also sets a standard for future annotation efforts in scientific document analysis.
\section{Variable Extraction Approaches}
\label{sec:baselines}
In our evaluation, we utilized diverse approaches, including traditional rule-based extraction models, popular LLMs with varying degrees of enhancement, and an optimized AI pipeline framework.
\subsection{Rule-based Information Extraction}
We developed a rule-based information extraction system\footnote{\url{https://github.com/ml4ai/skema/tree/main/skema/text_reading/scala}} using the Odin language \cite{valenzuela-escarcega-etal-2016-odins} to identify and extract variables mentioned in text alongside their associated definitions or descriptions and values associated with them. The rule-based system operates by matching patterns over the syntax of a sentence or phrase. Figure \ref{fig:rule} depicts an example rule. With the help of a linguist, we designed a set of rules to match different ways in which a concept or symbol (the variable) is defined (the description) in scientific papers. Similarly, another subset of rules to match numerical values and quantities associated to variables. Rule-based information extraction tools serve as complement to LLM and other deep-learning based approaches. They trade generalization and recall capabilities for higher precision and interpretability.

\begin{figure}[h]
    \begin{minted}[
    xleftmargin=10pt,
    linenos,
    escapeinside=%%,
    fontsize=\fontsize{5}{7}\selectfont,
    breaklines,
    breakanywhere
  ]{yaml}
- name: description_interpreted
  label: Description
  priority: ${priority}
  type: dependency
  example: "Beta can be interpreted as the effective contact rate."
  pattern: |
      trigger = [lemma="interpret"]
      description:Phrase = nmod_as
      variable:Identifier = nsubjpass
  \end{minted}
  \caption{Example of a pattern-matching rule system designed to detect variable descriptions. The word \texttt{interpreted} will anchor the pattern (line 8). Outgoing syntactic dependencies of types \texttt{nmod\_as} and \texttt{nsubjpass} to entities of types \texttt{Phrase} and \texttt{Identifier} link the rule's trigger to its description and variable arguments, respectively.}
  \label{fig:rule}
\end{figure}

\subsection{Vanilla LLM Extraction}
\label{sec:llms}
LLMs have demonstrated exceptional performance on a variety of semantic information extraction tasks. In our study, we established LLM baselines using a vanilla pipeline, in which each LLM was provided with only snippets of text on paper and tasked with extracting variable names, descriptions, and values. To optimize the effectiveness of our approach, we conducted extensive prompt engineering, iterating through more than ten rounds of refinement. These prompts were developed by a team of four PhD or postdoctoral researchers in computer science major, and the most effective prompt was selected for use in our evaluations. Figure \ref{fig:tool_prompt} illustrates the prompt template that was used in all LLM baselines. In this template, {\tt []} serves as a placeholder for the paper text, and the prompt specifies a structured format for the output, with default values provided for optional fields. {Additionally, we incorporate a few-shot prompting setup that provides language models with several examples within the prompt to enhance their performance.}

% \todo{Adding few shots experiments? Chunwei is working on the few shot experiment.}

% \begin{figure}[h]
% \centering
% \begin{tcolorbox}[
%     width=0.98\linewidth, % Adjust width to your preference
%     colback=lightorange, % Background color of the box
%     colframe=framecolor, % Frame color
%     sharp corners, % Option for sharp corners
%     boxrule=0.5pt, % Frame thickness
%     left=10pt, % Left margin within the box
%     right=10pt, % Right margin within the box
%     top=6pt, % Top margin within the box
%     bottom=6pt % Bottom margin within the box
% ]
%     \textbf{Prompt:} Please extract variable names and descriptions from the following paper text.\\
%     \textit{\textbf{[TEXT]}}\\
%     This text may contain variables or parameters and what they mean. If it does, list each of the variables on a separate line with the following attributes separated by "\textbar": \\
%     \textbf{name} \textbar{} \textbf{description} \textbar{} \textbf{numerical value}.\\
%     If the variable's value uses other variables or there is no value for the variable, output ``None'' for that variable value; do not hallucinate a variable value or variable description that does not exist in the text.
%     \\
% \textit{\textbf{[OPTIONAL\_EXAMPLES]}}
% \end{tcolorbox}
% \caption{Prompt for extracting variable names and descriptions from text. \todo{(We can remove this one and just keep the other prompt if short on space.)}}
% \label{fig:pure_prompt}
% \end{figure}

\subsection{Tool Enhanced LLM Extraction}
LLMs often share similar technical frameworks and have substantial overlap in their training datasets. This commonality can lead them to either overemphasize or overlook certain cases. To mitigate these biases and enhance extraction accuracy, it is beneficial to introduce additional perspectives. Therefore, beyond the standard evaluation using only the paper text, we have also incorporated outputs from a traditional model into our LLM evaluations. This approach is conceptually similar to the tool integration methods used in LangChain \cite{topsakal2023creating}; however, our objective is to generate a broader range of candidate options rather than to rely on the presumed high-quality outputs of these tools. As illustrated in Figure \ref{fig:tool_prompt}, these outputs are highlighted in blue font. The {\tt [TOOL EXTRACTION]} provided by the traditional model offers supplementary variable options for consideration. However, in cases of discrepancy, the original text is always prioritized to ensure the fidelity of the information extracted.

\begin{figure}[h]
\centering
\begin{tcolorbox}[
    width=0.98\linewidth, % Adjust width to your preference
    colback=lightorange, % Background color of the box
    colframe=framecolor, % Frame color
    sharp corners, % Option for sharp corners
    boxrule=0.5pt, % Frame thickness
    left=10pt, % Left margin within the box
    right=10pt, % Right margin within the box
    top=6pt, % Top margin within the box
    bottom=6pt % Bottom margin within the box
]
    \textbf{Prompt:} Please extract variable names and descriptions from the following paper text. \textcolor{highlightdiff}{You may refer to the provided tool extractions for your reference.} Here is some paper text: \\
    \textit{\textbf{[TEXT]}}\\
    This text may contain model related variables or parameters, their initial values and what they mean. If it does, list each of the variables on a separate line with the following attributes separated by "\textbar": \\
    \textbf{name} \textbar{} \textbf{description} \textbar{} \textbf{numerical value}.\\
    If the variable's value uses other variables or there is no value for the variable, output ``None'' for that variable value; do not hallucinate a variable value or variable description that does not exist in the text.\\
    \textcolor{brown}{\textit{\textbf{[OPTIONAL\_EXAMPLES]}}}\\
    \textcolor{blue}{Meanwhile, we get some variable extractions from another tool for your reference. These extractions may contain false positive or duplication cases. Please pay more attention to the true positive variables:} \\\textcolor{blue}{\textbf{\textit{[TOOL\_EXTRACTION]}}} \\\textcolor{blue}{Please try to extract variables on the original paper text first, then refer to the results from the tool extractions and see if you miss any variables. If you are not sure, please always check the original paper text.}
\end{tcolorbox}
\caption{Prompt templates for variable extraction using various setups. The black font indicates the prompt template for a standard LLM. The combination of black and brown fonts represents the template for few-shot prompting. The integration of black and blue fonts denotes the template enhanced by external tools.}
\label{fig:tool_prompt}
\end{figure}

\subsection{Optimized AI Pipeline Framework}
% Building on traditional models and LLMs, numerous systems have been proposed to automatically optimize AI-powered analytics and information extraction according to user preferences \cite{zheng2023efficiently,  chen2023seed, liu2024declarative,patel2024lotus,lin2024towards}. In this context, 
We also incorporate a system featuring a simple and declarative user interface.
Palimpzest is a system designed to streamline AI-powered analytics through declarative query processing \cite{liu2024declarative}. This system allows users to effortlessly specify analytical queries over unstructured data using a straightforward, Python-embedded declarative language. Users can define their desired data schema and attributes in natural language, enabling Palimpzest to automate complex optimization processes. This automation includes navigating various AI models, employing prompting techniques, and optimizing foundational models, thereby eliminating the need for the laborious tasks of manual pipeline tuning, model selection, and prompt engineering previously required when working with LLMs. By efficiently managing trade-offs between runtime, cost, and data quality, Palimpzest simplifies user interaction and significantly enhances the efficiency and cost-effectiveness of processing large-scale data. These capabilities position Palimpzest as a robust benchmark for evaluating the performance of AI-driven data processing systems in scientific and analytical contexts, ensuring substantial improvements in execution times and costs while maintaining or enhancing data quality.

\begin{figure}[t]
  % \centering
  \begin{minted}[
    xleftmargin=10pt,
    linenos,
    escapeinside=||,
    fontsize=\fontsize{6}{7}\selectfont,
    breaklines,
    breakanywhere
  ]{python}
import palimpzest as pz

class Variable(|\textcolor{codeblue}{pz.Schema}|):
    """ Represents a variable of a model in a scientific paper"""
    excerptid = |\textcolor{codeblue}{pz.Field}|(desc="The unique identifier for the excerpt", required=True)
    name = |\textcolor{codeblue}{pz.Field}|(desc="The label used for the scientific variable, like alpha or beta", required=True)
    description = |\textcolor{codeblue}{pz.Field}|(desc="A description of the variable", required=False)
    value = |\textcolor{codeblue}{pz.Field}|(desc="The value of the variable", required=False)

# define logical plan
excerpts = |\textcolor{codeblue}{pz.Dataset}|("snippets", schema=pz.TextFile)
output = excerpts.convert(Variable, desc="A variable used or introduced in the paper snippet", cardinality="oneToMany")

# user specified policy and execute plan
policy = |\textcolor{codeblue}{pz.MinimizeCostAtFixedQuality}|(min_quality=0.45)
results = |\textcolor{codeblue}{pz.Execute}|(excerpts, policy=policy)
  \end{minted}
  \caption{Palimpzest Code for Variable Extraction from Scientific Paper Snippets.}
  \label{fig:palimpzest-code}
\end{figure}

The Palimpzest code snippet shown in Figure \ref{fig:palimpzest-code} demonstrates a declarative approach to extracting variables from scientific paper excerpts. It defines the `Variable' class, which details a scientific variable found within the paper excerpt. This class includes fields for the variable's name, description, and value, with only the variable name being required. This setup efficiently captures the essential details needed for variable extraction, streamlining the process of transforming unstructured text into structured data suitable for further analysis.

The code then creates a dataset named "snippets" with the Palimpzest native `TextFile' schema and processes it to convert each snippet into instances of the `Variable' class, identifying variables mentioned in the text. This conversion cardinality `oneToMany' allows for multiple variables per snippet, reflecting the typical structure of scientific excerpts.

Finally, a user-specified policy (`MinimizeCostAtFixedQuality') is set to optimize the extraction process by minimizing operational costs while maintaining the quality of the extracted data above a predetermined threshold. The `Execute' function applies this policy to the dataset, demonstrating how Palimpzest simplifies complex data extraction tasks through its declarative programming model.
\section{Evaluation}
\subsection{Experimental Setup}
\begin{table*}
\centering
\caption{Average performance with GPT4 similarity evolution with ground-truth (bold font indicates the best over each setup).}
\resizebox{1.72\columnwidth}{!}{%
\label{tab:avg_gpt4_sim}
\begin{tabular}{@{}l|ccc|ccc|ccc@{}}
\toprule
\textbf{Model} & \multicolumn{3}{c|}{\textbf{Overall Performance}} & \multicolumn{3}{c|}{\textbf{Variable Descriptions}} & \multicolumn{3}{c}{\textbf{Variable Values}} \\
               & \textbf{Recall} & \textbf{Precision} & \textbf{F1} & \textbf{Recall} & \textbf{Precision} & \textbf{F1} & \textbf{Recall} & \textbf{Precision} & \textbf{F1} \\ 
\midrule
\texttt{pure\_GPT3.5T}   & 0.576 & 0.337 & 0.393 & 0.677 & 0.361 & 0.431 & 0.404 & 0.305 & 0.323 \\
\texttt{tool\_GPT3.5T}   & 0.568 & 0.307 & 0.369 $\downarrow$ & 0.647 & 0.318 & 0.396 & 0.429 & 0.281 & 0.306 \\
\texttt{3shot\_GPT3.5T}  & 0.543 & 0.412 & 0.437 $\uparrow$& 0.558 & 0.433 & 0.457 & 0.521 & 0.378 & 0.400 \\
\midrule
\texttt{pure\_GPT4T}     & 0.655 & 0.443 & \textbf{0.491} & 0.708 & 0.428 & 0.495 & 0.527 & 0.460 & 0.456 \\
\texttt{tool\_GPT4T}     & 0.662 & 0.451 & 0.500 $\uparrow$ & 0.711 & 0.440 & 0.506 & 0.559 & 0.478 & 0.476 \\
\texttt{3shot\_GPT4T}    & 0.645 & 0.502 & \textbf{0.535} $\uparrow$ & 0.650 & 0.471 & 0.514 & 0.629 & 0.557 & 0.553 \\
\midrule
\texttt{pure\_GPT4o}     & 0.647 & 0.424 & 0.480 & 0.708 & 0.438 & 0.504 & 0.513 & 0.382 & 0.408 \\
\texttt{tool\_GPT4o}     & 0.689 & 0.460 & 0.520 $\uparrow$ & 0.727 & 0.453 & 0.526& 0.589 & 0.468 & 0.483 \\
\texttt{3shot\_GPT4o}    & 0.499 & 0.360 & 0.389 $\downarrow$ & 0.486 & 0.376 & 0.395 & 0.525 & 0.369 & 0.397 \\
\midrule
\texttt{pure\_GPT4o-mini} & 0.619 & 0.376 & 0.437 & 0.693 & 0.410 & 0.479 & 0.499 & 0.325 & 0.360 \\
\texttt{tool\_GPT4o-mini} & 0.694 & 0.465 & \textbf{0.525} $\uparrow$ & 0.729 & 0.446 & 0.520 & 0.619 & 0.481 & 0.504 \\
\texttt{3shot\_GPT4o-mini} & 0.545 & 0.322 & 0.378 $\downarrow$& 0.578 & 0.370 & 0.426 & 0.475 & 0.231 & 0.282 \\
\midrule
\texttt{pure\_llama}     & 0.600 & 0.402 & 0.446 & 0.671 & 0.422 & 0.483 & 0.456 & 0.373 & 0.372 \\
\texttt{tool\_llama}     & 0.629 & 0.396 & 0.451 $\uparrow$ & 0.706 & 0.411 & 0.482 & 0.479 & 0.354 & 0.369 \\
\texttt{3shot\_llama}    & 0.488 & 0.181 & 0.244 $\downarrow$ & 0.550 & 0.204 & 0.279 & 0.402 & 0.139 & 0.180 \\
\midrule
\texttt{pure\_mistral}   & 0.572 & 0.234 & 0.301 & 0.661 & 0.220 & 0.302 & 0.404 & 0.285 & 0.310 \\
\texttt{tool\_mistral}   & 0.564 & 0.277 & 0.335 $\uparrow$ & 0.650 & 0.265 & 0.343 & 0.412 & 0.307 & 0.317 \\
\texttt{3shot\_mistral}   & 0.493 & 0.190 & 0.248 $\downarrow$ & 0.588 &0.191 &0.262 & 0.352 &0.194 &0.217 \\
\midrule
\texttt{rules}           & 0.392 & 0.317 & 0.320 & 0.447 & 0.352 & 0.358 & 0.299 & 0.244 & 0.245 \\
\midrule
\texttt{Palimpzest}      & 0.574 & 0.451 & 0.473 & 0.566 & 0.435 & 0.460 & 0.555 & 0.453 & 0.465 \\
% to remove from the aaai submission
structured\_GPT4o & 0.64 & 0.443 & 0.492 & 0.682 & 0.435 & 0.498 & 0.535 & 0.446 & 0.449 \\
structured\_GPT4o-mini & 0.658 & 0.424 & 0.484 & 0.689 & 0.406 & 0.476 & 0.593 & 0.442 & 0.473 \\
\bottomrule
\end{tabular}
}%
\end{table*}
We evaluate a variety of models to assess their performance on the variable extraction dataset. The traditional rule-based model is denoted as \texttt{rules}, and an optimized AI pipeline framework is referred to as \texttt{Palimpzest}. Additionally, we examine several advanced models from OpenAI, including \texttt{GPT3.5 Turbo}, \texttt{GPT4 Turbo}, \texttt{GPT4o}, and \texttt{GPT4o-mini}. We also test two locally served LLMs, \texttt{Llama-3-8B-Instruct} and \texttt{Mistral-7B-Instruct-v0.2}, which are integrated via vLLM model serving APIs. Each LLM is evaluated using a standard API call, indicated by the prefix \textit{pure\_}, and an enhanced version that incorporates outputs from the traditional model, indicated by the prefix \textit{tool\_}. The LLM temperature parameter is set to zero to ensure reproducibility.

We executed all baseline models using the prompts or configurations outlined in the previous section. The results are then aligned with the human-annotated ground truth, as illustrated in Figure \ref{fig:example_json}. This alignment is based on the input text chunk ID. Furthermore, we construct all possible candidate pairs by applying the Cartesian product to the sets of predicted extractions and ground truth, grouped by annotation type. This process resulted in a total of $330,558$ candidate pairs for evaluation. For each candidate pair, we employed a set of evaluation metrics to determine whether it qualified as a match.

To evaluate the F1 score in our study, we meticulously track the ground truth and prediction sets for each text chunk. During the evaluation process, when an evaluator confirms a match (though the criteria for a match may vary across different metrics), the index of the matched candidate pair is recorded in both the ground truth and prediction entries for that specific pair.  After evaluating all candidate pairs associated with a given text chunk, we calculate the recall as the ratio of entries with at least one match in the ground truth set. Similarly, precision is calculated as the ratio of entries with at least one match in the prediction set. The F1 score is then computed using the harmonic mean of precision and recall, providing a balanced measure of the model's accuracy in variable extraction tasks. {In cases where the evaluation focuses on specific tasks, such as variable descriptions or variable values extraction only, we count only the corresponding entries and disregard the others.}

\subsection{GPT-4 as a Similarity Evaluator}
We employed the GPT-4 turbo model to perform similarity evaluations, comparing its outputs with a ground-truth dataset to assess precision and accuracy across different tasks. Depending on whether the candidate pair being evaluated corresponds to "var\_desc" or "var\_val" (examples provided in Figure \ref{fig:example_json}), we use specific prompts as illustrated in Figure \ref{fig:gpt4_eval_prompts}. To ensure conciseness, we limit the output token length to one.

\begin{figure}[h]
\centering
\begin{tcolorbox}[
    width=0.98\linewidth,
    colback=lightorange,
    colframe=framecolor,
    sharp corners,
    boxrule=0.5pt,
    left=10pt,
    right=10pt,
    top=6pt,
    bottom=6pt,
    title=Prompt for Variable Description/Value:
]
    {You are a human evaluator. The following pair of text describes a variable and its description/value.}\\
    \textit{\textbf{[VAR\_DESC\_A]}}/\textit{\textbf{[VAR\_VAL\_A]}}\\
    \textit{\textbf{[VAR\_DESC\_B]}}/\textit{\textbf{[VAR\_VAL\_B]}}\\
    {Please check if they mean the same. Answer y or n.}
\end{tcolorbox}

% \begin{tcolorbox}[
%     width=0.98\linewidth,
%     colback=lightorange,
%     colframe=framecolor,
%     sharp corners,
%     boxrule=0.5pt,
%     left=10pt,
%     right=10pt,
%     top=6pt,
%     bottom=6pt,
%     title=Prompt for Variable Value:
% ]
%     {You are a human evaluator. The following pair of text describes a variable and its value.}\\
%     \textit{\textbf{[VAR\_VAL\_A]}}\\
%     \textit{\textbf{[VAR\_VAL\_B]}}\\
%     {Please check if they mean the same. Answer y or n.}
% \end{tcolorbox}
\caption{GPT4 Turbo prompt templates for evaluating the consistency of variable descriptions and values.}
\label{fig:gpt4_eval_prompts}
\end{figure}

% \begin{table}
% \centering
% \caption{Average performance with GPT4 similarity evolution with ground-truth}
% \label{tab:avg_gpt4_sim}
% \begin{tabular}{@{}lccc@{}}
% \toprule
% \textbf{Model}           & \textbf{Recall} & \textbf{Precision} & \textbf{F1}    \\ \midrule
% \texttt{pure\_llama}     & 0.600  & 0.402     & 0.446 \\
% \texttt{tool\_llama}     & 0.629  & 0.396     & 0.451 \\
% \texttt{pure\_mistral}   & 0.572  & 0.234     & 0.301 \\
% \texttt{tool\_mistral}   & 0.564  & 0.277     & 0.335 \\
% \texttt{pure\_GPT3.5T}   & 0.576  & 0.337     & 0.393 \\
% \texttt{tool\_GPT3.5T}   & 0.568  & 0.307     & 0.369 \\
% \texttt{pure\_GPT4T}     & 0.655  & 0.443     & 0.491 \\
% \texttt{tool\_GPT4T}     & 0.662  & 0.451     & 0.500 \\
% \texttt{pure\_GPT4o}     & 0.647  & 0.424     & 0.480 \\
% \texttt{tool\_GPT4o}     & 0.689  & 0.460     & 0.520 \\
% \texttt{pure\_GPT4o-mini} & 0.619 & 0.376     & 0.437 \\
% \texttt{tool\_GPT4o-mini} & 0.694 & 0.465     & 0.525 \\
% \texttt{rules}           & 0.392  & 0.317     & 0.320 \\
% \texttt{Palimpzest}      & 0.574  & 0.451     & 0.473 \\ \bottomrule
% \end{tabular}
% \end{table}

According to Table \ref{tab:avg_gpt4_sim}, no existing solution performs exceptionally well on the variable extraction task. However, the integration of rule-based approaches with LLMs has shown significant improvements. The best-performing baseline model achieved an F1 score of only $0.491$, while the integration with LLMs, particularly the GPT-4 variants, enhanced performance, achieving F1 scores as high as $0.525$. This represents a $20\%$ improvement over the setups using only LLMs, except for \texttt{GPT3.5T} where the integration did not yield a performance boost. Such integration highlights the complementary strengths of diverse methodologies: rule-based approaches provide additional variable extraction options from different perspectives, thereby enhancing the performance of LLMs.

Among the models tested, the tool-enhanced versions generally outperformed their pure counterparts, with \texttt{tool\_GPT4o-mini} achieving the highest F1 score of $0.525$. This indicates that the additional suggestions provided by tool extractions can effectively guide LLMs to achieve better performance. In contrast, the rule-based approach alone (\texttt{rules}) demonstrated lower effectiveness, with an F1 score of $0.320$, emphasizing the overall superior capability of LLM-based solutions in managing complex extraction tasks. 

{However, few-shot prompting does not consistently yield improved extraction results, as indicated by Table \ref{tab:avg_gpt4_sim}. Only GPT3.5T and GPT4T models showed improvement with the few-shot setting, while others experienced diminished performance. This variability could be attributed to the inherent complexity of the variable extraction task, where the diverse scenarios may not benefit significantly from a few additional examples. Moreover, the inclusion of more tokens in the prompt might dilute the attention mechanism, thereby worsening the results.}

\begin{table*}
\centering
\caption{Average token-level scores for variable descriptions and variable values.}
\resizebox{1.72\columnwidth}{!}{%
\label{tab:token-level}
\begin{tabular}{@{}l|ccc|ccc|ccc@{}}
\toprule
\textbf{Model}  & \multicolumn{3}{c}{\textbf{Overall Performance}} & \multicolumn{3}{c}{\textbf{Variable Descriptions}} & \multicolumn{3}{c}{\textbf{Variable Values}} \\
 & \textbf{Recall} & \textbf{Precision} & \textbf{F1} & \textbf{Recall} & \textbf{Precision} & \textbf{F1} & \textbf{Recall} & \textbf{Precision} & \textbf{F1} \\

\midrule
\texttt{pure\_GPT3.5T} & 0.622 & 0.578 & 0.552 & 0.730 & 0.663 & 0.645 & 0.458 & 0.449 & 0.410 \\
\texttt{tool\_GPT3.5T} & 0.551 & 0.527 & 0.505 $\downarrow$ & 0.636 & 0.625 & 0.599 & 0.421 & 0.379 & 0.362 \\
\texttt{3shot\_GPT3.5T} & 0.514 & 0.492 & 0.467 $\downarrow$ & 0.560 & 0.486 & 0.483 & 0.444 & 0.499 & 0.443 \\
\midrule
\texttt{pure\_GPT4T} & 0.661 & 0.587 & 0.571 & 0.770 & 0.670 & 0.666 & 0.496 & 0.460 & 0.428 \\
\texttt{tool\_GPT4T} & 0.667 & 0.638 & 0.610 $\uparrow$ & 0.771 & 0.712 & 0.695 & 0.508 & 0.527 & 0.482 \\
\texttt{3shot\_GPT4T} & 0.638 & 0.585 & 0.562 $\downarrow$ & 0.733 & 0.623 & 0.623 & 0.493 & 0.527 & 0.471 \\
\midrule
\texttt{pure\_GPT4o} & 0.664 & 0.621 & 0.595 & 0.759 & 0.688 & 0.673 & 0.521 & 0.520 & 0.477 \\
\texttt{tool\_GPT4o} & 0.667 & 0.620 & 0.599 $\uparrow$ & 0.768 & 0.686 & 0.681 & 0.513 & 0.521 & 0.475 \\
\texttt{3shot\_GPT4o} & 0.541 & 0.512 & 0.487 $\downarrow$ & 0.557 & 0.488 & 0.482 & 0.517 & 0.548 & 0.495 \\
\midrule
\texttt{pure\_GPT4o-mini} & 0.645 & 0.641 & \textbf{0.600} & 0.766 & 0.701 & 0.686 & 0.463 & 0.549 & 0.470 \\
\texttt{tool\_GPT4o-mini} & 0.659 & 0.691 & \textbf{0.640} $\uparrow$ & 0.771 & 0.773 & 0.738 & 0.489 & 0.567 & 0.490 \\
\texttt{3shot\_GPT4o-mini} & 0.644 & 0.589 & \textbf{0.564} $\downarrow$& 0.703 & 0.579 & 0.586 & 0.556 & 0.604 & 0.532 \\
\midrule
\texttt{pure\_llama} & 0.614 & 0.599 & 0.557 & 0.712 & 0.718 & 0.672 & 0.465 & 0.417 & 0.383 \\
\texttt{tool\_llama} & 0.621 & 0.630 & 0.585 $\uparrow$ & 0.723 & 0.780 & 0.716 & 0.466 & 0.401 & 0.385 \\
\texttt{3shot\_llama} & 0.554 & 0.553 & 0.511 $\downarrow$ & 0.614 & 0.607 & 0.573 & 0.461 & 0.470 & 0.417 \\
\midrule
\texttt{pure\_mistral} & 0.609 & 0.486 & 0.488 & 0.718 & 0.583 & 0.591 & 0.444 & 0.340 & 0.332 \\
\texttt{tool\_mistral} & 0.508 & 0.450 & 0.435 $\downarrow$ & 0.606 & 0.551 & 0.532 & 0.359 & 0.295 & 0.288 \\
\texttt{3shot\_mistral} & 0.564 & 0.489 & 0.482 $\downarrow$& 0.693 & 0.592 & 0.592 & 0.369 & 0.332 & 0.316 \\
\midrule
\texttt{rules} & 0.429 & 0.498 & 0.437 & 0.494 & 0.583 & 0.505 & 0.329 & 0.369 & 0.335 \\
\midrule
\texttt{Palimpzest} & 0.569 & 0.513 & 0.488 & 0.648 & 0.527 & 0.526 & 0.448 & 0.490 & 0.431 \\
\bottomrule
\end{tabular}%
}
\end{table*}

The Palimpzest system, utilizing \texttt{GPT-4o} as its conversion model, yielded results comparable to \texttt{pure\_GPT4o}, achieving an F1 score of $0.473$. By enforcing a strict format constraint, Palimpzest trades some recall for higher precision, offering a more reliable output without the need for extensive model selection and prompt engineering. This approach not only simplifies the extraction process but also enhances the usability and applicability of the system in practical scenarios, establishing a robust foundation for automatic model recovery and simulation.

{Additionally, we conducted a distinct quality assessment for both variable descriptions and variable values, with detailed results presented in Table \ref{tab:avg_gpt4_sim}. The observations mentioned above remain consistent across these evaluations. However, almost all baselines demonstrated better F1 scores on the variable descriptions task compared to their overall performance, with the exception of Palimpzest, which excelled in both cases in general but performed slightly better in the variable value extraction task.}

% Detailed violin plots illustrating the recall (\autoref{fig:recall_scores}), precision (\autoref{fig:precision_scores}), and F1 scores (\autoref{fig:f1_scores})  across 556 text chunks are available. \todo{Consideration may be given to removing these figures to save space for other metrics}.

% \begin{figure}[h]
%     \centering
%     \includegraphics[width=0.48\textwidth]{figures/f1_scores_violin.pdf}
%     \caption{F1 scores for each model. }
%     \label{fig:f1_scores}
% \end{figure}

% \begin{figure}[h]
%     \centering
%     \includegraphics[width=0.48\textwidth]{figures/precision_scores_violin.pdf}
%     \caption{Precision scores.}
%     \label{fig:precision_scores}
% \end{figure}

% \begin{figure}[h]
%     \centering
%     \includegraphics[width=0.48\textwidth]{figures/recall_scores_violin.pdf}
%     \caption{Recall scores.}
%     \label{fig:recall_scores}
% \end{figure}

\subsection{Token-based Evaluation}
In addition to the GPT-based evaluation, we examined the token-level precision, recall and F1 scores used for QA and other span prediction NLP tasks \cite{rajpurkar-etal-2016-squad}. Token-level scores account for the correct number of tokens predicted by each method, giving credit based on the proportion of tokens predicted correctly and penalizing for tokens predicted incorrectly. Table \ref{tab:token-level} shows the token level performance on the variable extraction dataset. The results don't diverge significantly from the GPT-based evaluation and consistently highlight the strength of LLM-based methods. Crucially, token-level scores rely solely on manual annotations, therefore any conclusions drawn from them are based only on the ground truth and not subject to any potential inaccuracies from a model-based evaluation.

% \subsection{BERT-score Evaluation}
% \todo{adding embedding similarity results}

% \subsection{Alpaca Eval \todo{(optional)}}

% \begin{figure}[h]
%     \centering
%     \includegraphics[width=0.6\textwidth]{figures/win_rates_heatmap-all.pdf}
%     \caption{Pairwise Alpaca evaluation.}
%     \label{fig:alpacaeval}
% \end{figure}

\subsection{Full Paper Context Extraction Evaluation}
{So far, the evaluations have been conducted on the dataset's passages, which inherently contain one or more annotations. The remaining contents of each article used to assemble the dataset have been ignored. Recently, LLMs started supporting longer context sizes for input, enough to frequently accommodate the full text of a scientific article.} We conducted variable extraction evaluations using the full text of each of the 22 articles. This approach limits the number of language models that can be used due to the token limits imposed by many LLMs. We present the results of a rule-based model, {GPT-3.5T (with chunking the long text into chunks within the model limit)}
% \todo{what does chunking means in this context? has it been explained before?}
, and GPT-4T. The overall performance is shown in Table \ref{tab:full_text}.

When dealing with the extensive context of a scientific paper, LLMs can struggle to maintain focus, often resulting in lower recall. In contrast, the \texttt{rules} and \texttt{pure\_GPT3.5T\_C} with chunking options manage to maintain relatively high recall. Overall, even in the context of lengthy texts, integrating tool outputs helps LLMs concentrate on the extraction task, leading to improved results.

\begin{table}
\centering
\caption{Overall Performance with Full Paper Text on Selected Models}
\label{tab:full_text}
\begin{tabular}{@{}lccc@{}}
\toprule
\textbf{Model} & \textbf{Recall} & \textbf{Precision} & \textbf{F1} \\
\midrule
\texttt{rules} & 0.701 & 0.101 & 0.172 \\
\texttt{pure\_GPT3.5T\_C} & 0.750 & 0.250 & 0.340 \\
\texttt{pure\_GPT4T} & 0.564 & 0.488 & 0.490 \\
\texttt{tool\_GPT4T} & 0.678 & 0.467 & \textbf{0.506} \\
\bottomrule
\end{tabular}
\end{table}
\section{Conclusion}

We have introduced a dataset for extracting variable descriptions and values from scientific literature, a crucial building block for the automated recovery of mathematical models from the literature. We conducted a battery of evaluations using different commercial and open-source LLMs, a rule-based information extraction system, and a declarative AI pipeline framework. In our experiments, we found that LLM-based methods tend to be the most effective methods to identify and extract variable descriptions and values; however, testing ensembles of rule-based and LLM-based information extractions working in tandem, boost the performance yield the best results most of the time. Considering that all the methods tested in this work did not use any form of supervised learning, there is ample room for improvement. In future work, multiple interesting avenues for research can be explored: Using semi-supervised and data-augmentation methods to augment the size of the dataset, and the use of supervised fine-tuning of encoder-based language models for generation and token prediction can improve the accuracy of the results.
\section{Acknowledgments}
We would like to express our gratitude for the support provided by the DARPA ASKEM Award HR00112220042. Additionally, we extend our appreciation to Patty Gahan and Robyn Kozierok from MITRE for their diligent annotation efforts.
%\clearpage

\bibliography{ref}

\end{document}